\documentclass[aps,showpacs,prb,reprint,superscriptaddress,groupedaddress]{revtex4-1}

\usepackage[colorlinks=true,linkcolor=blue,citecolor=blue,urlcolor=blue]{hyperref}

\usepackage{hyperref}
\usepackage{setspace} 
\usepackage{graphicx}
\usepackage{amsmath}
\usepackage{color}
\usepackage{amsmath}
\usepackage{amssymb}
\usepackage{verbatim}
\usepackage{latexsym}
\usepackage{enumerate} 
\usepackage{bm} 

\begin{document}

\title{Comparing electron-phonon coupling strength in diamond, silicon and silicon carbide: First-principles study}

\author{Bartomeu Monserrat}
\email{bm418@cam.ac.uk}
\affiliation{TCM Group, Cavendish Laboratory, University of Cambridge,
  J.\ J.\ Thomson Avenue, Cambridge CB3 0HE, United Kingdom}

\author{R.\ J.\ Needs}

\affiliation{TCM Group, Cavendish Laboratory, University of Cambridge,
  J.\ J.\ Thomson Avenue, Cambridge CB3 0HE, United Kingdom}

\date{\today}

\begin{abstract}
  Renormalization of the electronic band gap due to electron-phonon
  coupling in the tetrahedral semiconductors diamond, silicon and
  cubic silicon carbide is studied from first principles.  
There is a marked difference between the coupling of the vibrational state
  to the valence band maximum and to the conduction band minimum.  The
  strength of phonon coupling to the valence band maximum is similar between the three systems and is dominated by vibrations that change the bond length. The coupling strength to the conduction band minimum differs significantly in diamond, silicon carbide and silicon. In diamond, the coupling is dominated by six small pockets of vibrational states in the phonon Brillouin zone, that
  are ultimately responsible for the stronger electron-phonon coupling in this material.  Our results represent a first step towards the
  development of an \textit{a priori} understanding of electron-phonon
  coupling in semiconductors and insulators, that should aid the
  design of materials with tailored electron-phonon coupling
  properties.
\end{abstract}


\maketitle

\section{Introduction} 

Electron-phonon coupling is ubiquitous in condensed matter physics. It plays a central role in mediating the attractive interaction between electrons within the BCS theory of
conventional superconductivity,\cite{cooper_elph,BCS_theory} and it leads to the temperature dependence of electronic energy levels.\cite{RevModPhys.77.1173} Technological applications driven by
these effects include high-field
magnets,\cite{superconducting_magnets} photovoltaics, and light
emitting
diodes. 

The theoretical study of electron-phonon coupling in semiconductors was
established by the pioneering work of Allen, Heine, and
Cardona.\cite{0022-3719-9-12-013,PhysRevB.23.1495} Recently, their
ideas have been applied using accurate first-principles density
functional theory (DFT)
methods.\cite{0295-5075-10-6-011,PhysRevLett.105.265501,PhysRevB.87.144302,gonze_marini_elph}
DFT calculations not only give accurate predictions of the temperature
dependence of band gaps, but also allow us to investigate the
underlying microscopic physics.

The strength of electron-phonon coupling in semiconductors can be
gauged by the size of the correction to the band gap from the
zero-point (ZP) nuclear motion.
Experiments have focused on the effects of the isotopic mass
$m$,\cite{elph_C13,elph_Ge_isotope,RevModPhys.77.1173,Cardona20053}
and, as the amplitude of the atomic vibrations scales as $m^{-1/2}$, a
heavier isotope leads to weaker electron-phonon coupling. This
explanation is related to the qualitative considerations of Han and
Bester\cite{elph_Si_nano}, who compare the ZP gap correction for a
range of semiconductors, and find that the magnitude of the correction
is very roughly proportional 
to the ratio $\sqrt{\langle u^2\rangle} /a$, where $\sqrt{\langle
  u^2\rangle}$ is the mean atomic ZP displacement, and $a$ is the
lattice parameter. Electronic effects are considered by Cardona\cite{Cardona20053}, who argues that, as first row atoms
(like carbon) have no core $p$ electrons, the valence electrons
are strongly influenced by atomic vibrations, leading to stronger
electron-phonon coupling compared to systems with core $p$ electrons
(like silicon).

In this work we present a first-principles study of the strength of
electron-phonon coupling in the tetrahedrally bonded semiconductors
diamond (C), cubic silicon carbide (SiC), and silicon (Si). These
systems have the same crystal structure, and similar electronic band
structures and phonon dispersions. However, the effects of
electron-phonon coupling are significantly different between them, and
therefore they represent an interesting set of test systems.  Our
approach allows us to disentangle the vibrational and electronic
contributions and to evaluate their relative importance. A complex
picture emerges with different behaviour of the coupling of the
valence band minimum (VBM) and the conduction band maximum (CBM) to
the vibrational state. The electron-phonon coupling strength for the
VBM is similar in the three systems and dominated by vibrations that change the bond length between nearest neighbour atoms. The coupling strength for the CBM varies
significantly and in a non-trivial manner across the three systems 
studied, and accounts for the majority of the difference in the
coupling strengths of C, SiC, and Si. We find strong coupling in C which arises from the vibrational states in six small pockets in the phonon Brillouin zone (BZ).

This work represents a step towards a quantitative understanding of
the various factors influencing electron-phonon coupling in
semiconductors. This understanding could ultimately lead to the
development of strategies for designing new materials with tailored
electron-phonon coupling properties, a topic of great interest for
many technological applications.\cite{nature_phys_controlling_elph}


The paper is arranged as follows. In Sec.~\ref{sec:theory} we describe
the theoretical and computational methods used for calculating
electronic and vibrational properties, and their coupling. 
In Sec.~\ref{sec:results} we present our results for the various
contributions to the electron-phonon coupling strength in C, SiC, and
Si, and we summarize our findings in Sec.~\ref{sec:conclusions}. All
equations are given in Hartree atomic units, in which the Dirac
constant, the electronic charge and mass, and $4\pi$ times the
permittivity of free space are unity ($\hbar=|\mathrm{e}|=m_{\mathrm{e}}=4\pi
\epsilon_0=1$).

\section{Theoretical and computational frameworks} \label{sec:theory}


\subsection{Structural parameters} \label{subsec:dft}

We have studied C, SiC, and Si in their tetrahedrally bonded diamond
structures within plane-wave pseudopotential
DFT\cite*{PhysRev.136.B864,PhysRev.140.A1133} as implemented in the
{\sc{castep}} package.\cite{CASTEP} We have used the local density
approximation (LDA)\cite*{PhysRevLett.45.566,PhysRevB.23.5048}
functional and ultrasoft pseudopotentials\cite{PhysRevB.41.7892} for
C and Si. All energy differences between the different frozen phonon configurations used for the electron-phonon coupling calculations
are converged to better than $10^{-4}$ eV per unit cell, requiring a plane-wave energy cut-off of $800$ eV, 
and a Monkhorst-Pack\cite{PhysRevB.13.5188} BZ sampling grid
of density $2\pi\times0.03$ \AA$^{-1}$.

\begin{table}[b]
  \caption{Static LDA--DFT and experimental lattice parameters of C, 
    SiC, and Si in the diamond structure.} 
\label{tab:struct}
\begin{tabular}{lcc}
\hline
\hline
  & \hspace{0.2cm}  $a_{\mathrm{theor,static}}$ (\AA) & \hspace{0.2cm} $a_{\mathrm{exp}}$ (\AA) \\
\hline
C   & 3.529 & \hspace{0.2cm} 3.567\cite{diamond_a} \\
SiC & 4.318 & \hspace{0.2cm} 4.3581\cite{sic_a} \\
Si  & 5.394 & \hspace{0.2cm} 5.4298\cite{silicon_a} \\
\hline
\hline
\end{tabular}
\end{table}

We have relaxed the structures until the force on each atom is smaller than $10^{-4}$ eV/\AA\@ and the components of the stress tensor
are less than $10^{-2}$ GPa.  This leads to the lattice constants
shown in Table~\ref{tab:struct}, which are in good agreement with
experiment, and are used throughout. The LDA functional tends to overbind because it favours
uniform charge densities, as shown in Table~\ref{tab:struct}.

\subsection{Electronic structure}

The electronic configuration of C is $1s^22s^22p^2$, and when it forms
diamond the $2s$ and $2p$ electrons hybridize to form four $sp^3$ orbitals
which give rise to the tetrahedrally bonded structure. The electronic
configuration of Si is $1s^22s^22p^63s^23p^2$ and a similar
hybridization occurs, but with the $3s$ and $3p$ electrons. The
valence electrons in Si are farther away from the nucleus, which has
been invoked to explain the weaker electron-phonon coupling in Si
compared to diamond.\cite{Cardona20053}


\begin{figure}
\centering
\includegraphics[scale=0.78]{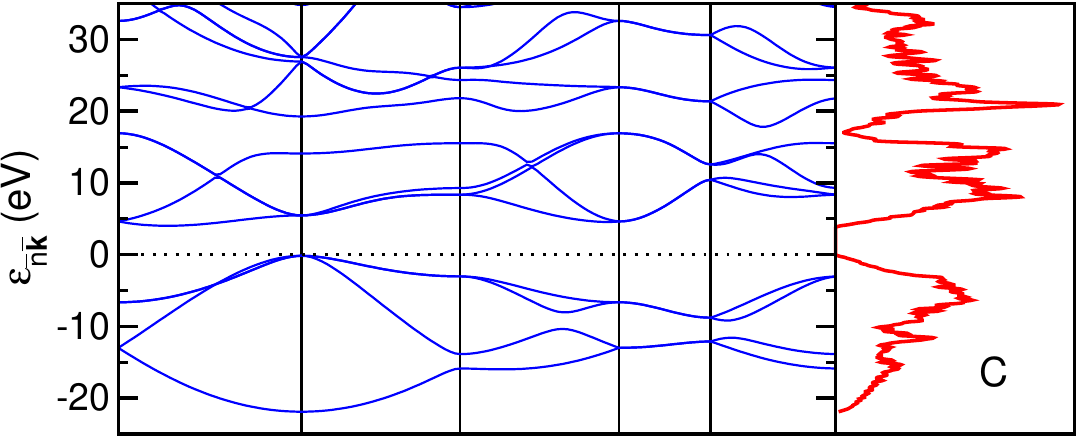}
\includegraphics[scale=0.78]{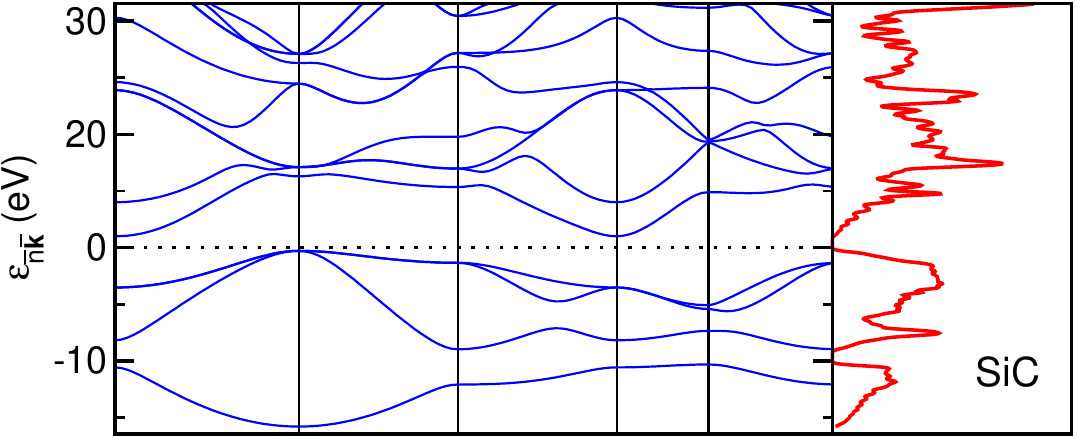}
\includegraphics[scale=0.78]{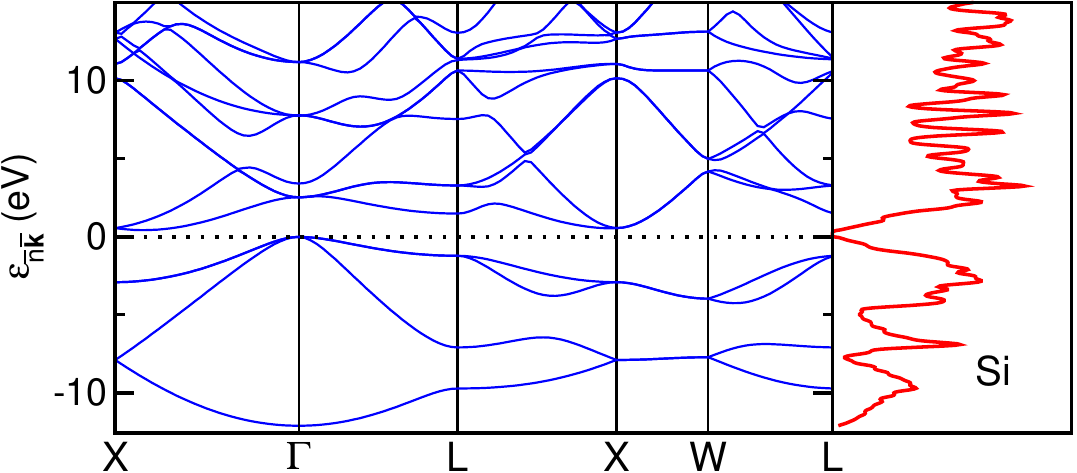}
\caption{(color online) Electronic band structures along symmetry
  lines of the first electron BZ (blue, left side), and electronic
  densities of states (red, right side) of C, SiC, and Si. The dotted
  line shows the VBM.  The wave vector axes have been
  scaled so that each plot has the same width.}
\label{fig:bs}
\end{figure}

In Fig.~\ref{fig:bs} we show the electronic band structures of C, Si,
and SiC along symmetry lines, and densities of electronic states. The
band structures are similar, with the VBM located at the
$\Gamma$-point and the CBM located along the symmetry line between
$\Gamma$ and $X$.

\subsection{Phonon dispersion} \label{subsec:ph_disp}

We have constructed the matrix of force constants from the forces on
the atoms calculated with finite atomic displacements of magnitude
$0.005$ \AA, averaging over positive and negative displacements.\cite{phonon_finite_displacement} We have then diagonalized the
dynamical matrix for the points along symmetry lines of the first
phonon BZ. Points in the phonon BZ are labelled by $\mathbf{k}$, and
branches by $n$. The force constants decay in real space as a function
of atomic separation, and we have found that for all three systems $5\times5\times5$ supercells
containing $250$ atoms leads to converged results for the vibrational energy.

\begin{figure}
\centering
\includegraphics[scale=0.35]{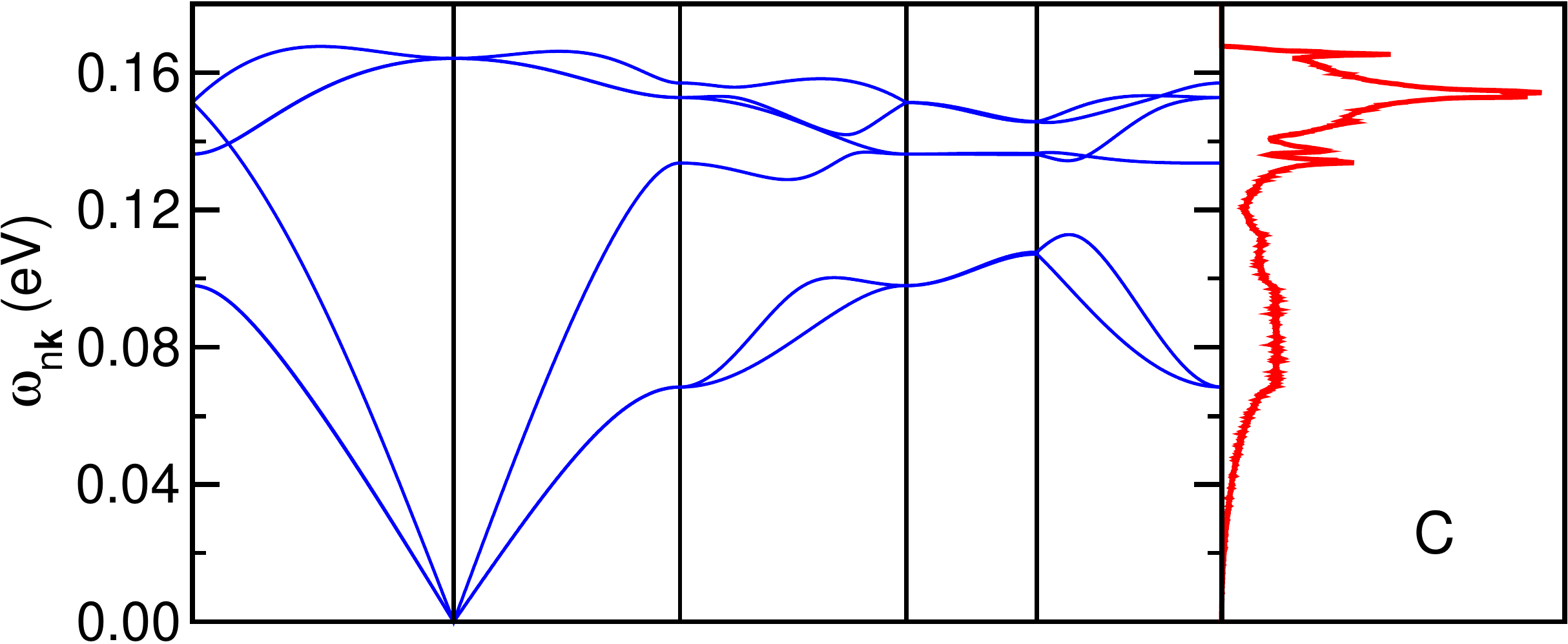}
\includegraphics[scale=0.35]{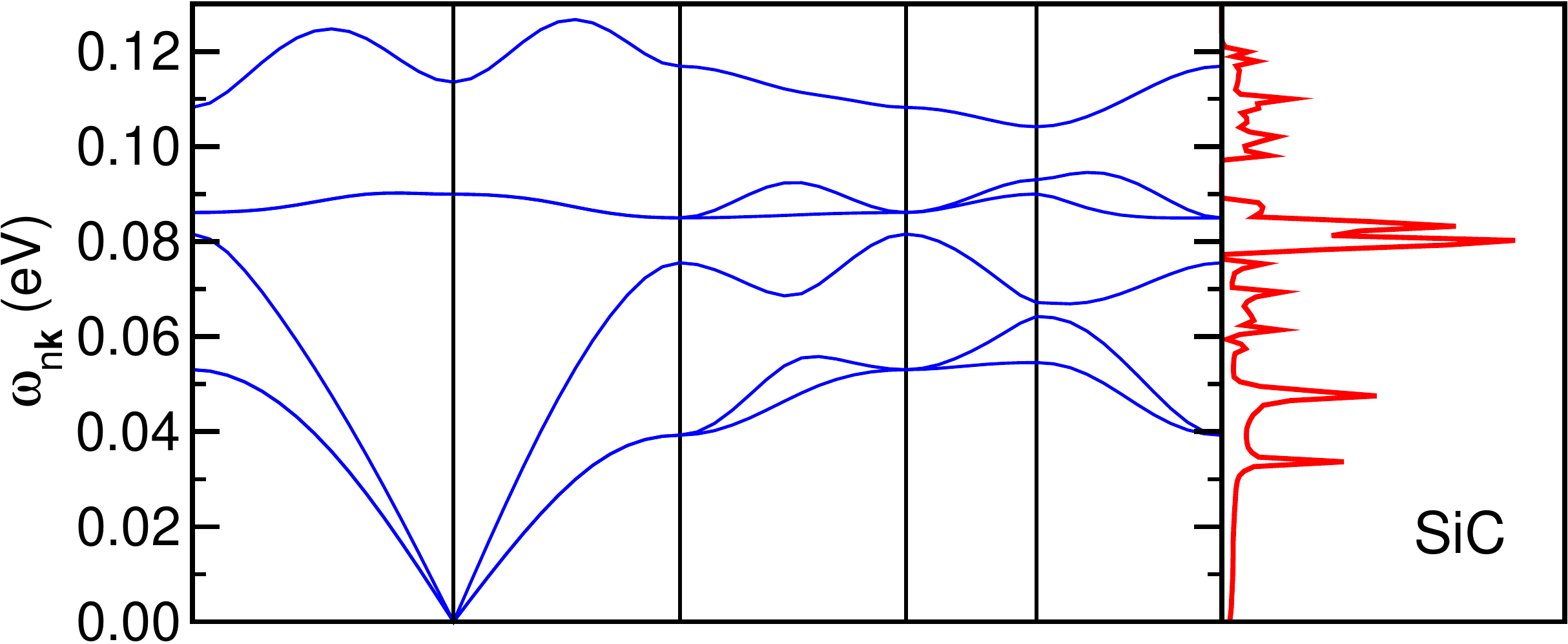}
\includegraphics[scale=0.35]{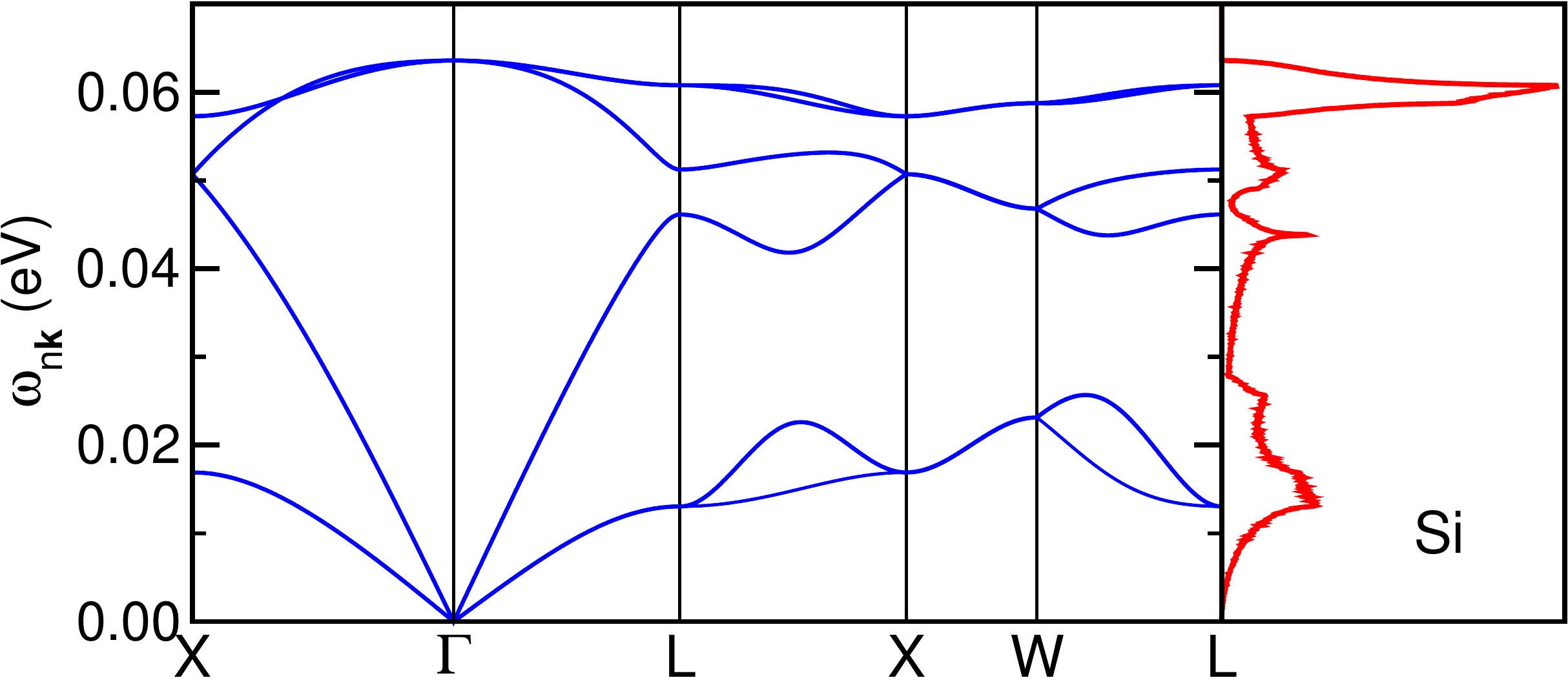}
\caption{(color online) Phonon dispersions along symmetry lines of the
  first phonon BZ (blue, left side), and phonon densities of states
  (red, right side) of C, SiC, and Si. The wave vectors axes have been
  scaled so that each plot has the same width.}
\label{fig:phdisp}
\end{figure}

In Fig.~\ref{fig:phdisp} we show the phonon dispersions and densities
of states of C, Si, and SiC. Considering SiC as an example, we have found 
that the highest energy phonon branch
corresponds to vibrations dominated by the motion of C atoms. The two
other optical branches correspond to antiphase vibrations of neighbouring Si and C atoms. The highest energy acoustic branch is dominated
by the motion of Si atoms, and the two lowest energy branches
correspond to long wave length vibrations. We note that the optical branches of SiC are subject to LO-TO splitting, calculated using density functional perturbation theory as implemented in the CASTEP code.\cite{lo_to_splitting_castep}

\begin{table}[b]
  \caption{Vibrational ZP energy per primitive unit cell of C, SiC, and Si 
    in the diamond structure at the theoretical equilibrium volumes.}
\label{tab:zpe}
\begin{tabular}{lc}
\hline
\hline
  & \hspace{0.2cm}  ZP energy (eV) \\
\hline
C   & 0.368 \\
SiC & 0.223 \\
Si  & 0.123 \\
\hline
\hline
\end{tabular}
\end{table}

The ZP energies per primitive unit cell of C, SiC, and Si are given in
Table~\ref{tab:zpe}. As expected, the lighter materials have larger ZP
energies. The ZP energies are converged to within $10^{-4}$ eV per
primitive unit cell with respect to the size of the supercell.

\subsection{Electron-phonon coupling}

The effect of electron-phonon coupling on the band gap of a
semiconductor can be calculated by considering the change in the
vibrational free energy arising from the promotion of an electron from
the valence to the conduction band,\cite{0295-5075-10-6-011} or by
calculating the change in the electronic bands due to the presence of
vibrations.\cite{0022-3719-9-12-013,PhysRevB.23.1495} These two
approaches can be shown to be equivalent, at least in lowest order
perturbation theory.\cite{ZPBAllen}

We use the second approach and calculate the zero temperature 
electron-phonon
correction to the electronic thermal (minimum) band gap
$E_{\mathrm{g}}$ as 
\begin{equation}
  \langle E_{\mathrm{g}}\rangle=\langle\Phi(\mathbf{q})|E_{\mathrm{g}}(\mathbf{q})|\Phi(\mathbf{q})\rangle, \label{eq:elph}
\end{equation}
where $|\Phi\rangle$ is the ground state vibrational wave function. 
We work within the harmonic
approximation, and therefore the vibrational wave function
$|\Phi\rangle$ is a Hartree product of simple harmonic
oscillator eigenstates for each vibrational mode, which are simple
Gaussian functions for the ground state.

The expression in Eq.~(\ref{eq:elph}) has been evaluated in the
literature by sampling $E_{\mathrm{g}}$ using path integral
methods\cite{PhysRevB.73.245202,ceperley_h_elph_coupling} or Monte
Carlo methods,\cite{giustino_nat_comm,helium} or by using some variant
of the expansion\cite{PhysRevB.87.144302,elph_Si_nano,analytic_arxiv}
\begin{equation}
E_{\mathrm{g}}(\mathbf{q})=\sum_{n,\mathbf{k}}a_{n\mathbf{k}}q_{n\mathbf{k}}^2, \label{eq:quadratic_expansion}
\end{equation}
where $q_{n\mathbf{k}}$ is the amplitude of the vibrational mode labeled
by $(n,\mathbf{k})$. Evaluating Eq.~(\ref{eq:elph}) using the
approximate expression of Eq.~(\ref{eq:quadratic_expansion}) is
similar to using the Allen-Heine-Cardona
theory, but the expression in Eq.~(\ref{eq:quadratic_expansion}) is more accurate because it includes the so-called non-diagonal Debye-Waller term that is missing in the Allen-Heine-Cardona theory.\cite{gonze_off_diagonal} 
We note that it excludes terms with higher powers of $q_{n\mathbf{k}}$
and coupling between different points in the BZ, whereas sampling
methods include all of these terms and should therefore be more
accurate. 
However, this expansion allows us to investigate the contribution from
each vibrational mode independently, which is obscured in a sampling
approach, and furthermore, the expansion in
Eq.~(\ref{eq:quadratic_expansion}) has been found to lead to very good
agreement with experiment for a range of materials, including
diamond.\cite{PhysRevLett.105.265501,PhysRevB.87.144302} Our
calculations of the ZP corrections to the band gaps of C, SiC, and Si
are shown in Table~\ref{tab:convergence} and compared with
experimental estimates of ZP band gap corrections where available.\footnote{In Ref.~\onlinecite{PhysRevB.87.144302} we reported a ZP band gap correction for diamond of $-0.462$~eV. Those calculations used a $3\times3\times3$ supercell, and should therefore be compared with the value of $-0.401$~eV reported in Table~\ref{tab:convergence}. The difference between these two values is due to the different treatment of the vibrational wave function and $E_{\mathrm{g}}(\mathbf{q})$. In Ref.~\onlinecite{PhysRevB.87.144302}, we used an anharmonic vibrational wave function and a principal axes expansion for $E_{\mathrm{g}}(\mathbf{q})$} 
The
experimental estimate of $-0.364$~eV for C is obtained from isotopic
data in Ref.~\onlinecite{Cardona20053}, and the estimate of
$-0.410$~eV from an extension of the typical Bose-Einstein oscillator
fit to the temperature dependence of the band gap in
Ref.~\onlinecite{analytic_arxiv}. 
In this work we calculate band gap
corrections using Eq.~(\ref{eq:quadratic_expansion}).

\begin{table}[b]
  \caption{ZP band gap corrections for C, SiC, and Si as a function of supercell size. 
    Experimental results are also shown where available.}
\label{tab:convergence}
\begin{tabular}{cccc}
\hline
\hline
BZ grid  &\hspace{0.2cm} C & SiC & Si \\
\hline
$3\times3\times3$ &\hspace{0.2cm} $-0.401$ eV  & $-0.110$ eV & $-0.053$ eV  \\
$4\times4\times4$ &\hspace{0.2cm} $-0.292$ eV  & $-0.089$ eV & $-0.052$ eV  \\
$5\times5\times5$ &\hspace{0.2cm} $-0.325$ eV  & $-0.109$ eV & $-0.060$ eV  \\
$6\times6\times6$ &\hspace{0.2cm} $-0.334$ eV  &  &   \\
\hline
 Exp. &\hspace{0.2cm} $-0.364$ eV\cite{Cardona20053} & & $-0.053$ eV\cite{Cardona20053} \\
 &\hspace{0.2cm} $-0.410$ eV\cite{analytic_arxiv} & & \\ 
\hline
\hline
\end{tabular}
\end{table}

The strategy we follow in Sec.~\ref{sec:results} is to evaluate the
couplings $a_{n\mathbf{k}}$ for each vibrational mode labelled by
$(n,\mathbf{k})$. This allows us to determine
which vibrational modes contribute to the overall electron-phonon
correction to the band gap. We work within the Born-Oppenheimer
approximation, which implies that the couplings $a_{n\mathbf{k}}$ are
independent of temperature. Therefore, without loss of generality, we
focus on the ZP correction to the band gap.

We use the harmonic wave functions obtained as described in
Sec.~\ref{subsec:ph_disp} and calculate the couplings
$a_{n\mathbf{k}}$ by performing frozen-phonon calculations for the
$\mathbf{k}$-points in the irreducible phonon BZ. The frozen phonon
calculation for the vibrational mode labelled by $(n,\mathbf{k})$ is
performed at a vibrational amplitude of magnitude about $\sqrt{\langle q^2_{n\mathbf{k}} \rangle}$, and we have averaged over positive and negative displacements.  Our
results are for supercells constructed from $6\times6\times6$ primitive
unit cells in the case of C, and from $5\times5\times5$
primitive unit cells for SiC and Si, unless otherwise stated. 
The ZP band gap correction for C converges slowly with respect to the
size of the supercell (see Table~\ref{tab:convergence}). 
This is a well-known feature in
C,\cite{gonze_marini_elph} and the slow convergence with respect to
phonon BZ sampling will prove to be intimately related to the
electron-phonon coupling strength in C, as discussed in
Sec.~\ref{subsec:cbm}. 

We also note that the use of DFT for calculating electron-phonon
induced band gap corrections is appropriate, although LDA-DFT is not
accurate for the calculation of the absolute value of band gaps.
The usual band gap underestimation of
standard DFT approximations affects all frozen-phonon configurations
in a very similar manner, and therefore it is expected to cancel in the
calculation of the band gap correction as it is the difference between
the static band gap $E_{\mathrm{g}}$ and the renormalized band gap
$\langle E_{\mathrm{g}}\rangle$. 
This is supported by the numerical
results of Giustino and co-workers in
Ref.~\onlinecite{PhysRevLett.105.265501}.

\section{Results}  \label{sec:results}

The ZP band gap correction for C is several hundreds of meVs, while in
Si it is about $6$ times smaller. In this section we investigate
the underlying microscopic properties that give rise to the ZP band
gap corrections.

\subsection{Microscopic description of the strength of electron-phonon coupling} \label{subsec:strength}

\begin{figure}
\centering
\includegraphics[scale=0.34]{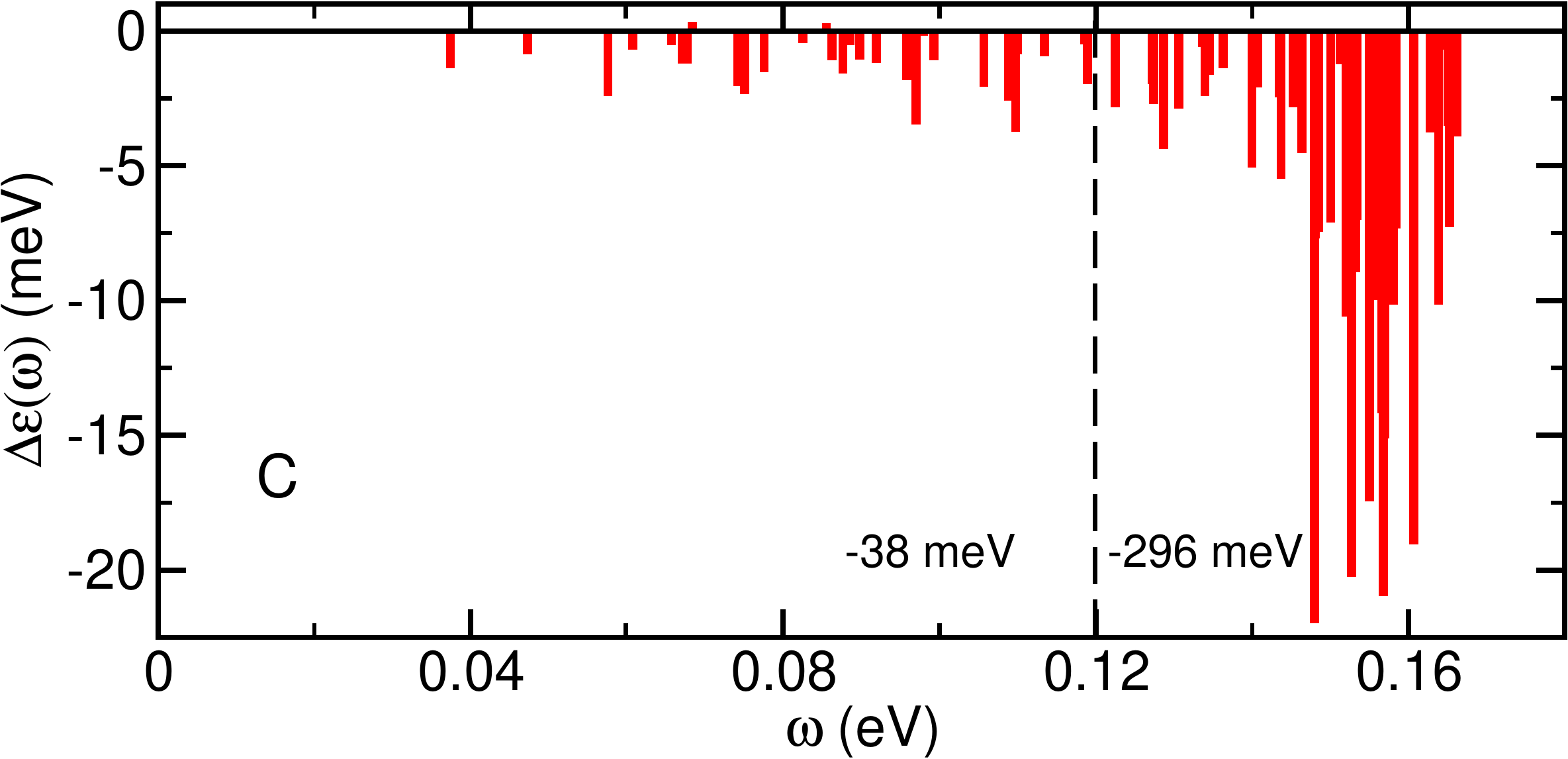}

\vspace{0.2cm}

\includegraphics[scale=0.34]{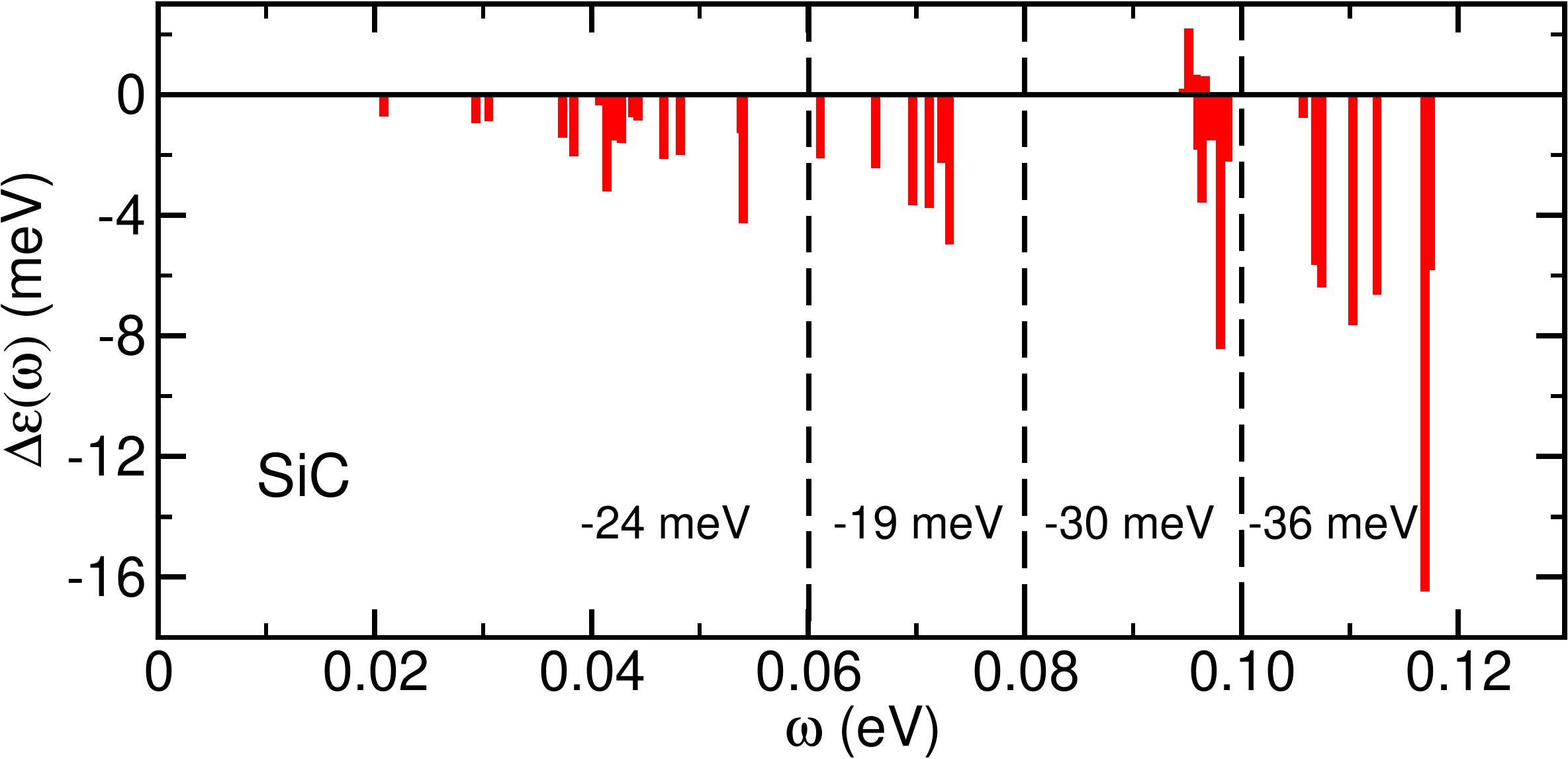}

\vspace{0.2cm}

\includegraphics[scale=0.34]{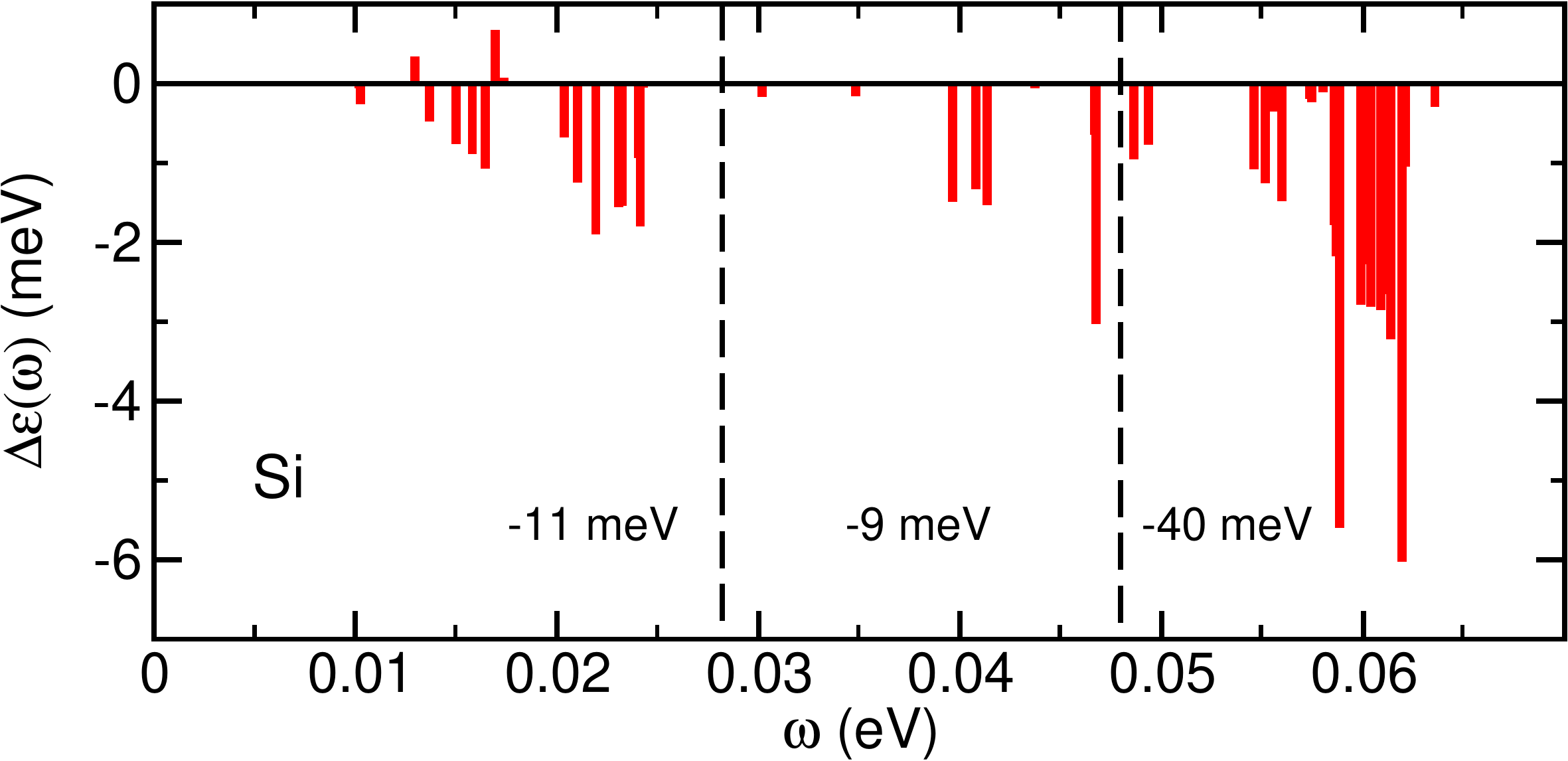}
\caption{(color online) ZP band gap corrections $\Delta\epsilon(\omega)$ as a function of the
  frequency $\omega$ of the vibrational modes for C, SiC, and Si. The vertical dashed lines separate
  intervals of the phonon branches with different vibrational characters
  (see text for details). The numbers reported for each of these intervals correspond to the integrated correction to the band gap for that interval.} 
\label{fig:vbm_and_cbm}
\end{figure}

\begin{figure*}
\centering
\includegraphics[scale=1.50]{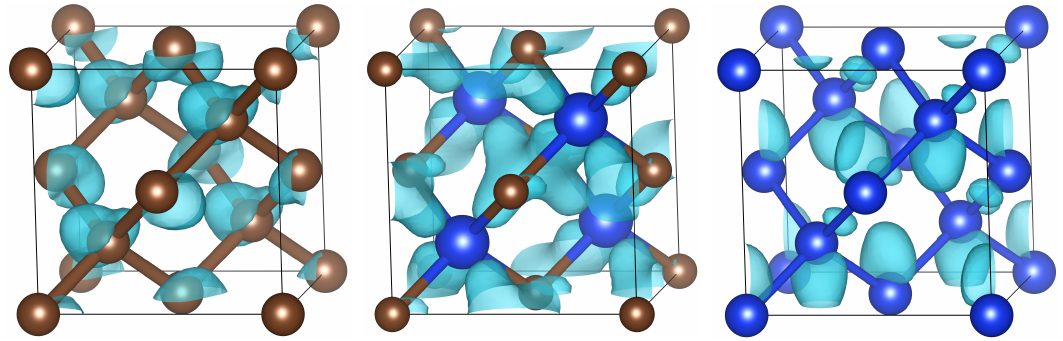}
\includegraphics[scale=1.50]{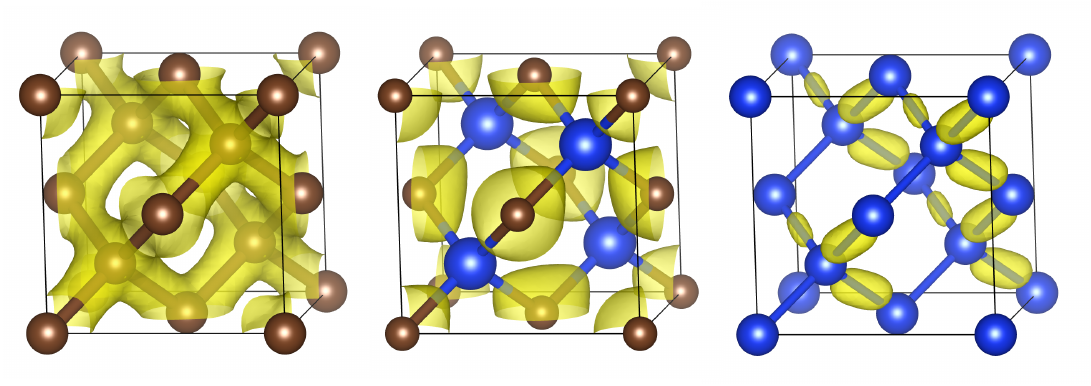}
\caption{(color online) Charge density isosurfaces of the electronic
  state corresponding to the CBM (top three) and VBM (bottom three) for diamond (left), silicon carbide
  (center), and silicon (right). The structures are not to
  scale.} 
\label{fig:structures_vbm}
\end{figure*}

In Fig.~\ref{fig:vbm_and_cbm} we show the electron-phonon contribution $\Delta\epsilon(\omega)$
to the band gap correction as a function of the
harmonic vibrational mode frequency $\omega$ for C, SiC, and Si. These
results correspond to a supercell containing $432$ atoms ($1296$ vibrational modes) for C, and $250$ atoms ($750$
vibrational modes) for SiC and Si, which is equivalent to a phonon BZ sampling using $6\times6\times6$ and $5\times5\times5$ grids, respectively.

For C, the phonon density of states (see
Fig.~\ref{fig:phdisp}) can be divided into low- and high-energy
branches, with a division at about $0.12$ eV. The high-energy
branches correspond to vibrational modes in which nearest-neighbour
atoms vibrate out of phase. For SiC, as discussed in
Sec.~\ref{subsec:ph_disp}, the vibrational modes can be divided into
the motion of C atoms, optical vibrations, the motion of Si atoms, and
acoustic vibrations. These divisions correspond to energies of
$0.06$ eV, $0.08$ eV, and $0.10$ eV. In the case of Si, we have
divided the vibrational modes into three classes delimited by energies
of $0.028$ eV and $0.048$ eV. These divisions are shown as vertical
dashed lines in Fig.~\ref{fig:vbm_and_cbm}. In each case
we have integrated the electron-phonon contribution to the change in
the gap for the specified vibrational frequency intervals $(\omega_1,\omega_2)$ as $\int_{\omega_1}^{\omega_2}\Delta\epsilon(\omega)d\omega$, which are also given in Fig.~\ref{fig:vbm_and_cbm}.

In all three systems the dominant contribution to the correction to
the band gap arises from the highest energy vibrational modes, which
correspond to optical phonons. In C the correction due to the optical modes clearly dominates, but for SiC and Si the low energy modes make an important contribution as well.

\subsection{Valence band maximum}

\begin{figure}
\centering
\includegraphics[scale=1.05]{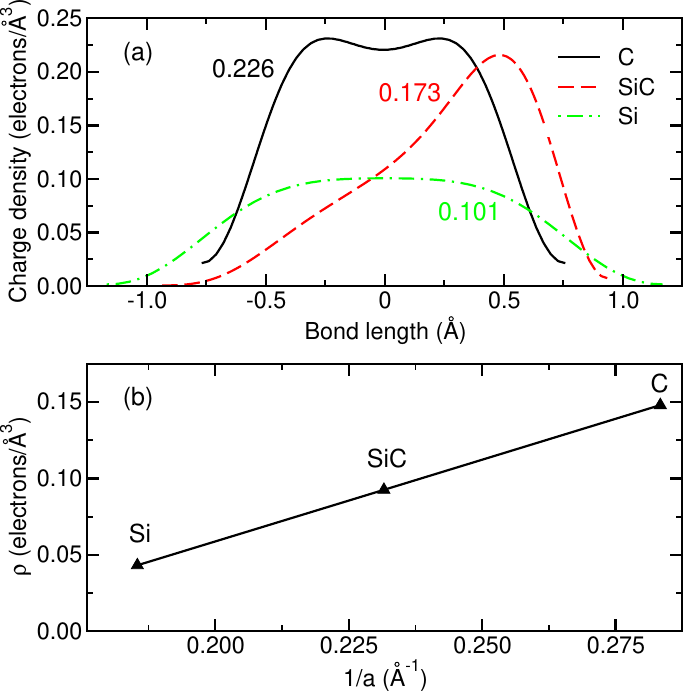}
\caption{(color online) (a) Charge density of the electronic state
  corresponding to the VBM along a bond 
  of C (solid black line), SiC (dashed red line), and Si
  (dashed-dotted green line). In each case we also give the
  integrated charge density along the bonds, averaged over the four
  bonds of the tetrahedral structure. For SiC, the C atom is at the
  right of the figure, where the charge density is larger. (b) The
  average bond density $\rho$ as a function of the inverse lattice constant $a^{-1}$.}
\label{fig:bond_density}
\end{figure}

In the bottom row of Fig.~\ref{fig:structures_vbm} we show the charge density of the VBM
for C, SiC, and Si, which is concentrated around the bonds for C and
Si. SiC is an ionic material, and the valence
charge density is therefore attracted towards the more electronegative C nuclei, although
still oriented along the bond direction. The electronic charge density
along the bonds is shown in the upper diagram of
Fig.~\ref{fig:bond_density}. The location of the charge density
agrees with the observation from Fig.~\ref{fig:vbm_and_cbm} that in each material the
dominant contribution to the band gap correction arises from optical modes,
as these vibrations change the length of the
interatomic bonds in which the charge density associated with the VBM resides. 

The ZP band gap correction of C and Si over the phonon BZ is shown in Fig.~\ref{fig:BZ_contour} in the $2$-dimensional slice defined by $\mathbf{c}^{\ast}=\textbf{0}$. The vectors ($\mathbf{a}^{\ast},\mathbf{b}^{\ast},\mathbf{c}^{\ast}$) are the reciprocal lattice vectors.
The vibrations that couple strongly to the VBM charge density by changing the bond length form a shell around the $\Gamma$-point in the phonon BZ, indicated in Fig.~\ref{fig:BZ_contour} by the red solid circles.

The lower diagram of Fig.~\ref{fig:bond_density} demonstrates that
the average charge density $\rho$ along the bond obeys $\rho\propto a^{-1}$, where $a$ is the lattice parameter. 
This simple dependence of the charge density associated with the VBM, together with the very similar coupling to the VBM between C and Si shown in Fig.~\ref{fig:BZ_contour}, cannot explain the strong non-linearity found by Han and Bester in the band gap correction as a function of lattice parameter $a$.\cite{elph_Si_nano} 

\subsection{Conduction band minimum} \label{subsec:cbm}

The top row of Fig.~\ref{fig:structures_vbm} also shows the charge density of the CBM for
C, SiC, and Si. The charge density is localized around 
the atomic sites, in contrast to the bond localized charge density of the VBM. 

\begin{figure}
\centering
\includegraphics[scale=0.90]{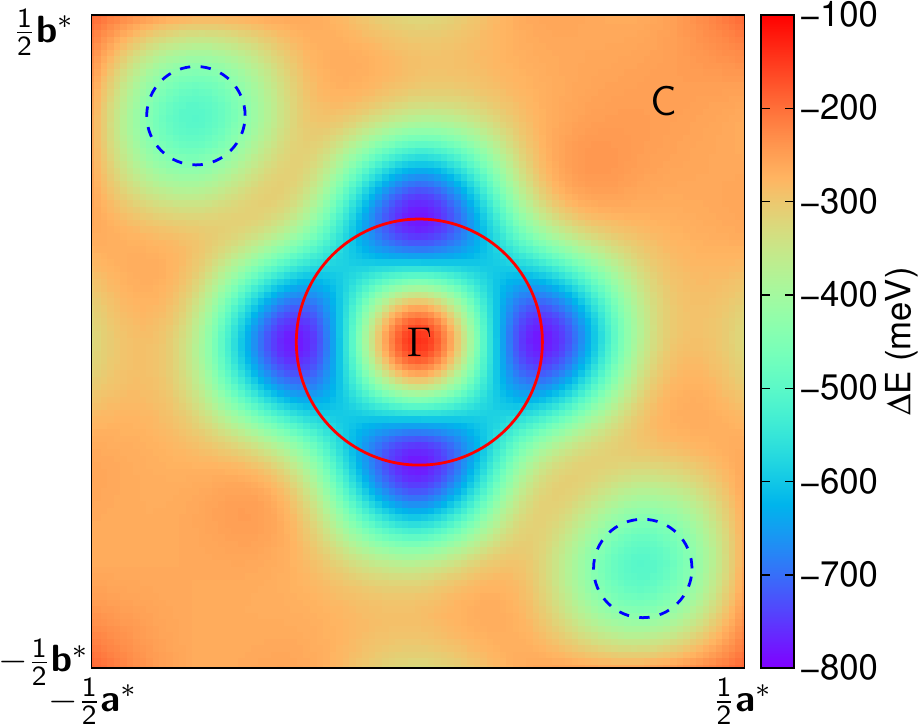}
\includegraphics[scale=0.90]{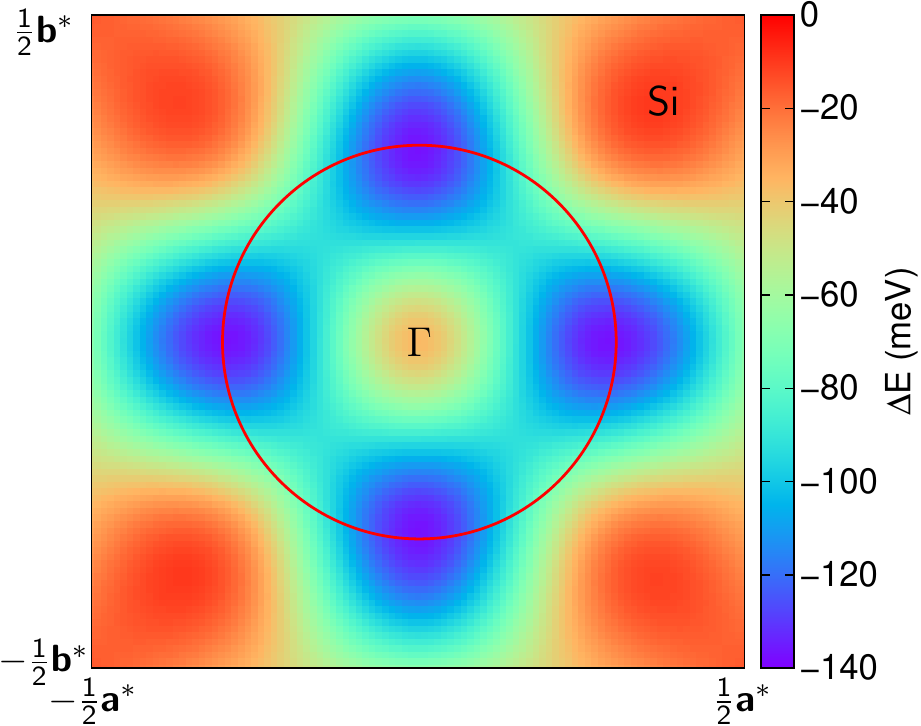}
\caption{(color online) ZP correction to the band gap of C and Si over $2$-dimensional slices of the phonon BZ defined by $\mathbf{c}^{\ast}=\mathbf{0}$. For each $\mathbf{k}$-point we have summed over the six phonon branches present. The red solid circles indicate the region of strong electron-phonon coupling of the VBM, and the dashed blue circles indicate the region of strong coupling to the CBM. The overall ZP correction to the band gap is $-334$~meV for C and $-60$~meV for Si.}
\label{fig:BZ_contour}
\end{figure}

Two major features can be seen in the ZP band gap correction for C over the phonon BZ shown in Fig.~\ref{fig:BZ_contour}. The first is the shell around the $\Gamma$-point (red solid circle), that is a consequence of the coupling of optical-like vibrations that change the bond lengths and couple strongly with the charge density associated with the VBM as discussed above. This feature is present in both C and Si, as the VBM charge density is very similar between these systems. However, in C (and only in C), there is a further strong variation, indicated by blue dashed circles in Fig.~\ref{fig:BZ_contour}. This second strong variation in C is
dominated by modes in the neighbourhood of the reciprocal space point
($\mathbf{a}^{\ast}/3$,$-\mathbf{b}^{\ast}/3$,$\mathbf{0}$) and symmetry-related points in the phonon BZ, where
the ZP correction to the band gap is large. There are a total of six such spherical-like pockets in the phonon BZ, of which we show a cross-section of two in Fig.~\ref{fig:BZ_contour}. 

\begin{figure}
\centering
\includegraphics[scale=0.90]{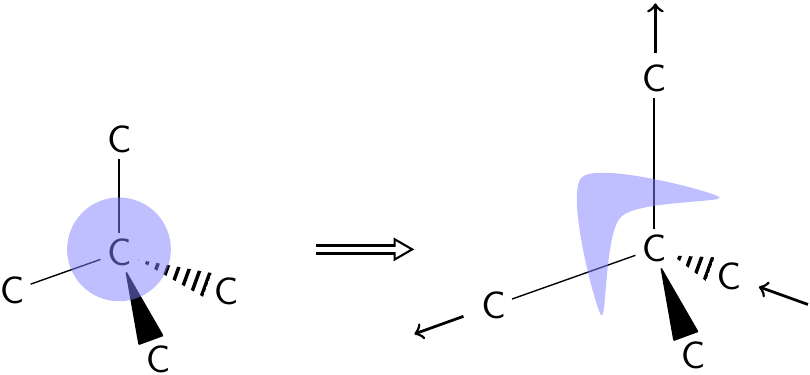}


\caption{(color online) Schematic of the bond distortion in the C tetrahedra for the vibration at the reciprocal-space point ($\mathbf{a}^{\ast}/3$,$-\mathbf{b}^{\ast}/3$,$\mathbf{0}$) in the phonon BZ. The charge density (blue) accumulates near the two stretched bonds due to its antibonding nature.}
\label{fig:schematic_bond}
\end{figure}

The charge density corresponding to the bottom of the conduction bands of C, SiC, and Si is antibonding. When the structures are distorted along vibrational modes, the charge density changes in
such a way as to preserve the antibonding nature.
For the mode at the reciprocal space point ($\mathbf{a}^{\ast}/3$,$-\mathbf{b}^{\ast}/3$,$\mathbf{0}$), the charge density in C is distorted significantly from the
equilibrium charge density, whereas the corresponding distortion in Si
is smaller. The vibrational mode has a wavelength of $3$
primitive unit cells, and the local tetrahedral structure around a
particular C atom is modified such that one bond remains at its
equilibrium length, one (or two) bonds lengthen, and two
(or one) bonds shorten. The charge density is then distorted so that it is
concentrated along the directions of the longer bonds, as would be
expected of an antibonding state (see Fig.~\ref{fig:schematic_bond}). In Si, the charge density responds
in a similar manner, but the changes are much smaller, leading to 
significantly weaker electron-phonon coupling than in C.

It is worth pointing out that it is also this difference between C and Si that leads to the slower convergence of the band
gap correction of C with the density of BZ sampling (see
Table~\ref{tab:convergence}).

\begin{figure}
\centering
\includegraphics[scale=0.80]{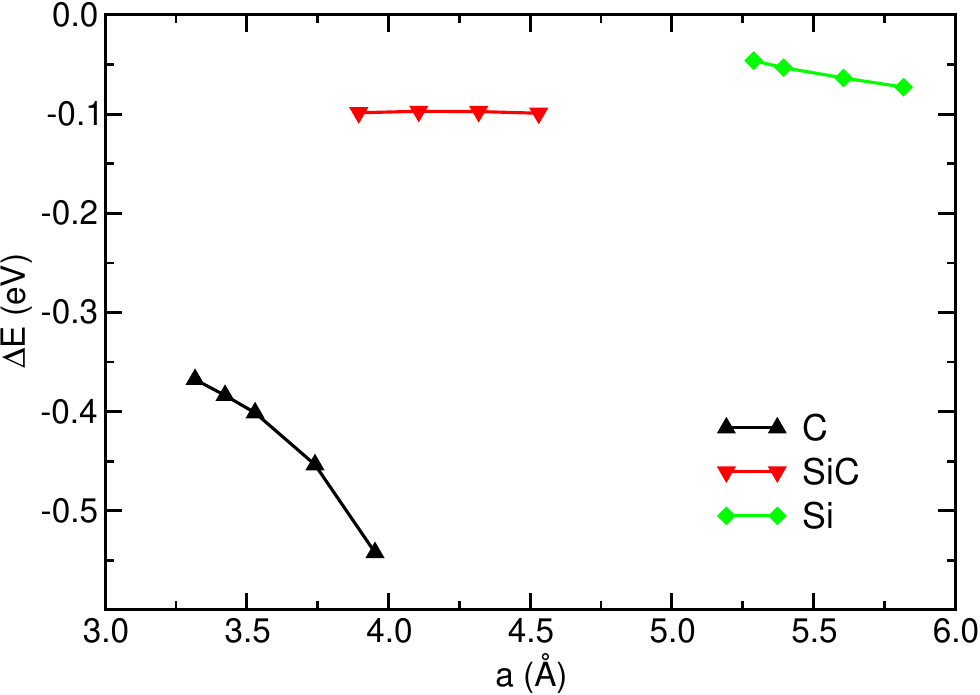}
\caption{(color online) ZP band gap correction for C (black up triangles), SiC (red down triangles), and Si (green diamonds) as a function of lattice parameter $a$. The solid lines are a guide to the eye.}
\label{fig:pressure}
\end{figure}

In order to investigate the origin of the difference in responses of
the charge densities of C and Si, we consider the correction to the band gap
due to electron-phonon coupling as a function of lattice parameter.
The changes in lattice constant shown in Fig.~\ref{fig:pressure} are
equivalent to the application of an external isotropic compression
(favoring smaller lattice constants) or dilation (favoring a larger
lattice constant). These calculations use a supercell containing
$3\times3\times3$ primitive unit cells and $54$ atoms, which is
sufficient for our purposes, as the $3\times3\times3$ supercell leads to converged results for Si, and includes the points in the phonon BZ of diamond that couple strongly to the electronic CBM (see Fig.~\ref{fig:BZ_contour}). The electron-phonon coupling strength in C
increases with lattice constant $a$. This increase is dominated by the six pockets in the phonon BZ that correspond to coupling to the CBM,
and that are still located around the point
($\mathbf{a}^{\ast}/3$,$-\mathbf{b}^{\ast}/3$,$\mathbf{0}$), and correspond to similar atomic displacements for the structures at different $a$. The stronger coupling with
increasing $a$ can be understood by a complementary increase in
$\sqrt{\langle u^2\rangle}$ for the modes dominating the coupling, which means that
$\sqrt{\langle u^2\rangle}/a$ is an increasing function of $a$. Therefore, the ratio $\sqrt{\langle u^2\rangle}/a$
 plays an important role in the strength of electron-phonon coupling within C. The different
responses of C and Si are caused by the shorter bond lengths
in C, where the atomic displacements represent a larger fraction
of the bond length, and induce a stronger charge distortion
to preserve the antibonding nature of the CBM.

It is interesting to note that the results shown in Fig.~\ref{fig:pressure} have implications for the electron-phonon coupling correction to the band gap of C as a function of temperature. Increasing temperature leads to thermal expansion, which increases the lattice parameter $a$. The strong dependence of the band gap correction on $a$ in C suggests that a study of the temperature dependence of the band gap should include calculations of the correction as a function of volume.

\subsection{Discussion}

The strength of electron-phonon coupling as reflected by the ZP
correction to the VBM is similar in the three tetrahedral
semiconductors considered. This emerges from the coupling of
the optical vibrational modes to the VBM charge density, which is
localized along the bonds.

In contrast, the picture for the ZP correction to the CBM is more
complex. In all three systems, for a given vibrational amplitude the charge density distorts in order to
preserve the antibonding nature of the CBM, but while this distortion is moderate in SiC and Si, it is
very strong for C, especially in some regions of the phonon BZ. This is related to the larger value of
the ratio $\sqrt{\langle u^2\rangle}/a$ in C compared to SiC and Si.

It is useful to reconsider earlier work in view of our findings. The
arguments put forward previously in order to explain the difference in
electron-phonon coupling strength between C and Si can be classified
as vibrational arguments and electronic arguments:

\noindent (1) \textit{Vibrational arguments}.  Han and
Bester\cite{elph_Si_nano} observe in a range of semiconductors that
the electron-phonon coupling strength is an increasing function of
$\sqrt{\langle u^2\rangle}/a$, albeit in a very non-linear fashion. 
We have found that this non-linearity arises from the different vibrational couplings to the CBM in the three systems, which has a much larger amplitude in C for the
modes around six localized pockets of the phonon BZ. These vibrational
modes lead to very strong electron-phonon coupling. This suggests that
the non-linear behaviour observed by Han and Bester is dominated by
the coupling of the CBM to the vibrational states of the system.

\noindent (2) \textit{Electronic arguments}. Cardona\cite{Cardona20053} has argued
that the different strengths of electron-phonon coupling in C and Si
are due to the presence of core $p$-states in Si that screen the atomic
potential from the valence electrons more effectively than in C. However, the charge densities of the VBM and CBM differ
significantly from atomic densities, and it is the distortion of the
bonds rather than the motion of the nuclei with respect to an
atomic-like charge density that drives electron-phonon coupling.  It is interesting to note that the
larger charge density in the bonds of C is due to the denser
structure of C. The core $p$ states in Si therefore \textit{indirectly}
reduce the strength of electron-phonon coupling, but by making Si less dense than C, rather than by directly
affecting the distortions of the charge density via
vibrations.

The core $p$ states of Si in SiC lead to localization of the charge
density along the bonds near the carbon atoms (see the top diagram of
Fig.~\ref{fig:bond_density}).  However, this is due to the presence of
\textit{both} carbon and silicon atoms. In Fig.~\ref{fig:vbm_and_cbm}
we observe that the vibrational modes dominated by carbon motion
(correction of $-36$ meV, for $\omega>0.10$ eV) lead to
a larger correction to the band gap than those dominated by the motion of silicon (correction of $-19$ meV, for 
$0.06<\omega<0.08$ eV). However, if we renormalize the carbon motion
value to take into account the enhancement of the vibrational motion of
carbon atoms due to their smaller mass, we find $\sqrt{m_{\mathrm{C}}/m_{\mathrm{Si}}}\times(-36)=\sqrt{12/28}\times(-36)=-24$~meV, which is only slightly larger than
the silicon-motion-dominated correction ($-19$ meV), and is not enough to explain
the difference in strength between the ZP band gap
corrections of C and Si. Furthermore, the marginal difference is due to the charge
density being more strongly localized near the carbon atom in the
bond, but this only arises because \textit{both} carbon and silicon are
present, and it would not occur in pure C or Si. This discussion
suggests that the simple electronic argument of Cardona \cite{Cardona20053}
cannot account for the different electron-phonon coupling strengths
in C and Si.

\section{Conclusions} \label{sec:conclusions}

We have investigated electronic and vibrational properties of C, SiC,
and Si using first principles DFT calculations. We have discussed the
electron-phonon coupling induced correction of the electronic
thermal band gaps of these semiconductors, focusing on the underlying
microscopic properties. This has allowed us to disentangle the various
contributions to the band gap corrections arising from coupling
between the vibrations and electronic states. We have found 
very different behaviour in the coupling of the vibrational state to
the VBM and CBM. The ZP correction to the VBM can be described in a
similar manner for all three systems. In contrast, for the CBM there are
marked differences between C and Si/SiC, with the stronger coupling in C
arising from a small number of vibrational modes in six small quasi-spherical
pockets in the phonon BZ. This stronger coupling arises from the larger ratio
$\sqrt{\langle u^2\rangle}/a$ in C. The stronger coupling of the CBM to the vibrational motion in C is responsible for the overall larger band gap correction than in Si/SiC.

This is, as far as we are aware, the first quantitative investigation
of the underlying processes which determine the strength of
electron-phonon coupling in semiconductors. Our results for 
C, SiC, and Si should not be extended to
other types of semiconducting materials without careful consideration,
as they might show different behaviour. Similar studies of other
classes of semiconductors could increase our understanding
of the relative importance of the contributions to electron-phonon
coupling in systems with a band gap, and this knowledge could
prove valuable in the design of new materials with tailored
properties. 

Finally, we note that a similar approach could be taken to investigate
the underlying microscopic properties of the coupling between the
vibrational state of a solid and physical quantities other than the electronic band gap.

\begin{acknowledgements}
  We thank Michael Rutter for implementing the calculation of the
  charge density along a line in the {\sc check2xsf} program.  Financial support
  was provided by the Engineering and Physical Sciences Research
  Council (UK). The calculations were performed on the Cambridge High
  Performance Computing Service facility and the ARCHER UK National Supercomputing Service (http://www.archer.ac.uk), for which access was obtained via the UKCP consortium, EP/K013564/1.
\end{acknowledgements}

\bibliography{/Users/bartomeumonserrat/Documents/research/phd/thesis/anharmonic}

\end{document}